\documentclass[aps,prd,twocolumn,groupedaddress]{revtex4}
\usepackage{graphicx}
\usepackage{dcolumn}
\usepackage{bm}
\begin{document}

\title{Shear viscosity and spectral function of the quark matter}

\author{Masaharu Iwasaki, Hiromasa Ohnishi$^1$, and Takahiko Fukutome}
\email{miwasaki@cc.kochi-u.ac.jp}
\affiliation{Department of Physics, Kochi University, Kochi 780-8520, Japan \\
$^{1}$Institute of Materials Structure Science, KEK, Tsukuba 305-0801, Japan}
\email{ohni@post.kek.jp}

\date{\today}

\begin{abstract}
We discuss the shear viscosity of the quark matter by using the Kubo-Mori formula. It is found that the shear viscosity is expressed in terms of the quark spectral function. If the spectral function is approximated by a modified Bright-Wigner type, the viscosity decreases as the width of the spectral function increases. We also discuss dependence of the shear viscosity on the temperature and the density.
\end{abstract}


\maketitle

The existence of the quark gluon plasma (QGP), which is predicted by Quantum chromodynamics, has not been discovered in Nature. In order to produce such a new state of matter, the experimental investigation started at the Relativistic Heavy Ion Collider (RHIC). The data at RHIC, however, seems to reveal some unexpected properties of the high density matter produced in the experiment \cite{IA05}-\cite{KA05}. Namely it could be explained by a fluid model with small viscosity; it is almost perfect fluid. This fact encourages many researchers to calculate the transport coefficients of the quark matter \cite{HK85D}-\cite{HD05}. Moreover many studies by the lattice QCD have been in progress \cite{KW87}-\cite{NS05}.

It is the purpose of this note to calculate the shear viscosity of the quark matter. There are two kinds of particles in the QGP: quarks and gluons. In this paper, we take up only the quark sector so that the Nambu-Jona-Lasinio (NJL) model is available for the quark matter. As for calculation method of the viscosity, we use the correlation function method which is called Kubo-Mori formula \cite{K57}-\cite{Z74}.

According to the Kubo-Mori formula \cite{M62}, the shear viscosity $\eta(T)$ at temperature $T$ is given by
\begin{equation}
\eta(\omega) =\frac{1}{T}\int_{0}^{\infty}\mathrm{d}t\mathrm{e}^{i\omega
 t}\int\mathrm{d}{\bf r}(J_{xy}(\mathbf{r},t),J_{xy}(0,0)),
\end{equation}
where $J_{xy}$ is the $x,y$ component of the energy-momentum tensor of the quark matter. The correlation function in the right-hand side is defined by
\begin{equation}
(A,B) \equiv
 \beta^{-1}\int_{0}^{\beta}\mathrm{d}\lambda\langle\mathrm{e}^{\lambda
 H}A\mathrm{e}^{-\lambda H}B\rangle,
\end{equation}%
where $A$ and $B$ are operators of any physical quantities and $H$ denotes our Hamiltonian. The bracket, $\langle A \rangle = {\rm Tr}(A\mathrm{e}^{-\beta H})/{\rm Tr}\mathrm{e}^{-\beta H}$, means the thermal average at temperature $T$ ($\beta\equiv 1/T$). Using partial integration in the right-hand side of Eq.(1), the viscosity can be transformed into 
\begin{equation}
\eta(\omega) =\frac{i}{\omega}[\Pi^\mathrm{R}(\omega)-\Pi^\mathrm{R}(0)].
\end{equation}%
Here $\Pi^R (\omega)$ is a retarded Green's function defined by
\begin{equation}
\Pi^R (\omega) =-i\int_{0}^{\infty}\mathrm{d}t\mathrm{e}^{i\omega
 t}\int\mathrm{d}{\bf r}\langle[J_{xy}(\mathbf{r},t),J_{xy}(0,0)]\rangle,
\end{equation}%
where [ , ] in the integrand denotes the commutation relation. Noting that $(\Pi^R (\omega))^{*}=\Pi^R (-\omega)$, the (static) viscosity is reduced to
\begin{equation}
\eta\equiv \eta(\omega=0)=\left.-\frac{d}{d\omega}{\rm Im}\Pi^R (\omega)\right|_{\omega=+0}.
\end{equation}%
In order to calculate the above $\Pi^R (\omega)$, it is convenient to transform into the imaginary time (Matsubara) formalism. We introduce the following correlation function,
\begin{equation}
\Pi(\omega_n) =-\int_{0}^{\beta}\mathrm{d}\tau\mathrm{e}^{{-i\omega_{n}\tau}}\int\mathrm{d}{\bf r}\langle T_\tau(J_{xy}(\mathbf{r},\tau) J_{xy}(0,0))\rangle,
\end{equation}%
where the Matsubara frequency is represented by $\omega_{n}=2\pi nT$ and $T_{\tau}$ means the (imaginary) time ordering operator. As is well known, the retarded Green's function $\Pi^R (\omega)$ is obtained by the analytic continuation: $\Pi^R (\omega) =\left.\Pi(\omega_{n})\right|_{i\omega_{n}=\omega+i\delta}$.

We take the NJL model for the quark matter in this paper \cite{HK94}. Then the canonical energy-momentum tensor is read as
\begin{equation}
J_{xy} =\frac{i}{2}[\bar\psi\gamma^2\partial^1\psi-\partial^1\bar\psi\gamma^2\psi],
\end{equation}
where $\psi$ is the field operator for quarks. If this expression is substituted into Eq.(6), the correlation function $\Pi$ is written as 
\begin{equation}
\Pi(\omega_n)
 =\frac{1}{\beta}\sum_l\int\frac{\mathrm{d}{\bf p}}{(2\pi)^3}{p_x}^2{\rm Tr}[{\gamma^2}G({\bf p},\omega_l+\omega_n){\gamma^2}G({\bf p},\omega_l)],
\end{equation}%
where $G({\bf p},\omega)$ is the full propagator for the quark. This expression corresponds to the Feynman diagram drawn in Fig.1. 
\begin{figure}
 \includegraphics[width=\linewidth]{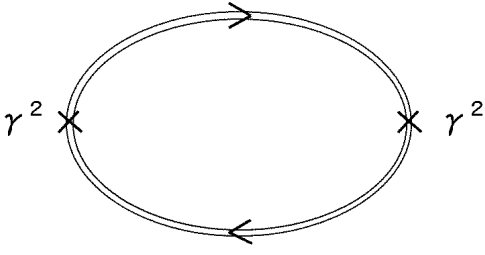}
 \caption{\label{fig:epsart}The Feynman diagram for the correlation function $\Pi$ in Eq.(8).}
\end{figure}
Of course there are many higher-order diagrams involved in this process, such as the ring (bubble) diagrams like the Nambu-Goldstone (pion) mode \cite{HK85}. But in the present case, such terms do not contribute to the $\Pi$ because the trace factor in the $\Pi$ is suppressed due to the presence of the $\gamma^2$. Hereafter we are restricted ourselves to the simple diagram of Fig.1.

Now let us follow the procedure taken for the calculation of the electrical conductivity in Ref.\cite{M90}. We consider the spectral representation for the full propagator, which is written as
\begin{equation}
G_{\alpha\beta}({\bf p},\omega_l) =\int_{-\infty}^\infty
 \frac{\mathrm{d}\varepsilon}{2\pi} \frac{\rho_{\alpha\beta}({\bf p},\varepsilon)}{i\omega_l-\varepsilon}.
\end{equation}
We substitute this expression into the correlation function (Eq.(8)) and the summation over Matsubara frequency is replaced by the contour integral:
\begin{eqnarray}
S&\equiv &T\sum_l {\rm Tr}[{\gamma^2}G({\bf p},\omega_l+\omega_n){\gamma^2}G({\bf p},\omega_l)] \nonumber \\
&=&-\int_C \frac{\mathrm{d}z}{2\pi i} n(z) {\rm Tr}[G({\bf p},z)\gamma^2 G({\bf p},z+i\omega_n)\gamma^2],
\end{eqnarray}
where $n(z)=(1+e^{\beta z})^{-1}$ is the Fermi distribution function. The countour $C$ is divided into three pieces as shown in Fig.2, because the integrand has branch cuts on the two lines $z=\varepsilon$ and $z=\varepsilon-i\omega_n$ where $\varepsilon$ is real (Note that poles of $n(z)$ do not lie on the two branch cuts).
\begin{figure}
 \includegraphics[width=\linewidth]{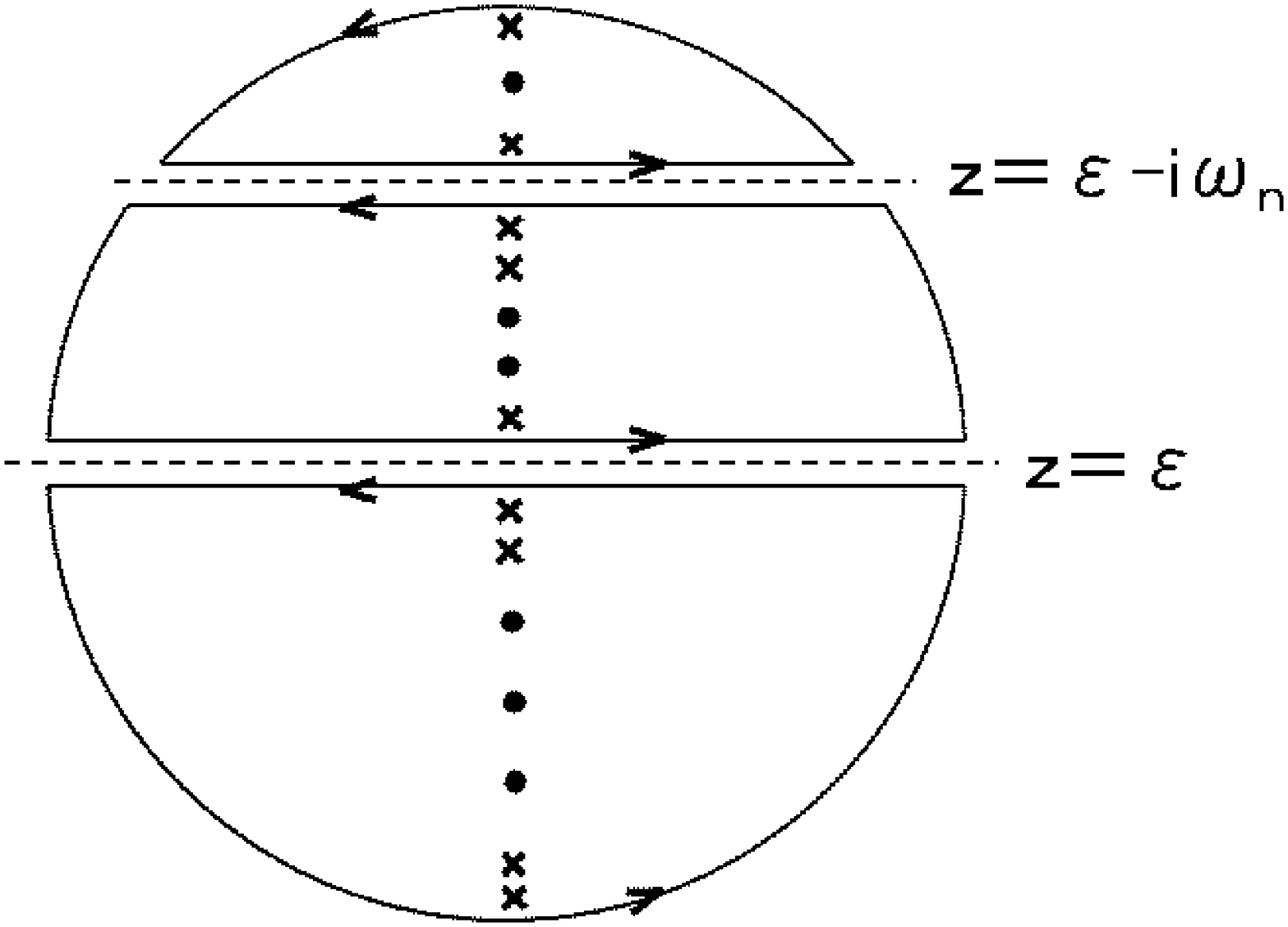}
 \caption{\label{fig:epsart}The countour for the calculating the integral of $z$ in Eq.(10).}
\end{figure}
Since the integral along the large circle vanishes, it is rewritten as
\begin{eqnarray}
S&=&-\int_{-\infty}^{\infty}\frac{\mathrm{d}\varepsilon}{2\pi
  i}n(\varepsilon){\rm Tr} [G(\varepsilon+i\delta)\gamma^2 G(\varepsilon+i\omega_n)\gamma^2 \nonumber \\
&& 
  -G(\varepsilon-i\delta)\gamma^2 G(\varepsilon+i\omega_n)\gamma^2  +G(\varepsilon-i\omega_n)\gamma^2 \nonumber \\
&& 
   \times G(\varepsilon+i\delta)\gamma^2  -G(\varepsilon-i\omega_n)\gamma^2 G(\varepsilon-i\delta)\gamma^2],
\end{eqnarray}%
where $\delta$ is an infinitesimal positive number introduced in order to avoid the branch cuts. Noting the relation, $G(\varepsilon+i\delta)-G(\varepsilon-i\delta)=-i\rho(\varepsilon)$, the above equation is expressed as follows:
\begin{equation}
S = \int_{-\infty}^{\infty} \frac{\mathrm{d}\varepsilon}{2\pi} n(\varepsilon) {\rm Tr}[(G(\varepsilon+i\omega_n)+G(\varepsilon-i\omega_n))\gamma^2\rho(\varepsilon)
\gamma^2].
\end{equation}%
If the analytic continuation $i\omega_{n}\longrightarrow \omega+i\delta$ is carried out, the imaginary part of the above equation is
\begin{equation}
\mbox{Im}S = \int_{-\infty}^{\infty} \frac{\mathrm{d}\varepsilon}{2\pi}
 \frac{1}{2} (n(\varepsilon+\omega)-n(\varepsilon)) {\rm Tr}[\rho(\varepsilon+\omega)\gamma^2\rho(\varepsilon)\gamma^2],
\end{equation}%
because ${\rm Tr}[\rho\gamma^{2}\rho\gamma^{2}]$ is real. As a result we obtain the shear viscosity expressed by the quark spectral function,
\begin{equation}
\eta = -\frac{1}{2}\int_{-\infty}^{\infty}
 \frac{\mathrm{d}\varepsilon}{2\pi}
 \int\frac{\mathrm{d}{\bf p}}{(2\pi)^3}{p_x}^2\frac{\partial n}{\partial
 \varepsilon}
  {\rm Tr}[\rho(\varepsilon)\gamma^2\rho(\varepsilon)\gamma^2].
\end{equation}
This equation means that the calculation of the shear viscosity is reduced to that of the spectral function. The appearance of ${\partial n}/{\partial \varepsilon}$ represents the Pauli blocking effect (See Eq.(18)).

Next stage is to discuss the quark spectral function. We take the following simple parameterization for the retarded (advanced) Green's function,
\begin{eqnarray}
G^R ({\bf p},\omega) &=& \frac{1}{p\cdot \gamma-M+i\mathrm{sgn}(p_0)\Gamma}, \nonumber \\
G^A ({\bf p},\omega) &=& \frac{1}{p\cdot \gamma-M-i\mathrm{sgn}(p_0)\Gamma},
\end{eqnarray}%
with the definition of $p_{0}\equiv \omega+\mu$ ($\mu$: the chemical potential). Here $M$ and $\Gamma$ represent the effective mass and the width of the quark respectively. These quantities are not determined and regarded as parameters in this paper. From these equations, we get a spectral function of modified Bright-Wigner type,
\begin{equation}
 \rho({\bf p},\omega)
  =\frac{1}{i}(G^A ({\bf p},\omega)-G^R ({\bf p},\omega))=\frac{2\Gamma\mathrm{sgn}(p_0)}{(p\cdot \gamma-M)^2+\Gamma^2}.
\end{equation}
It should be noted that there exists a sum rule $\int \rho({\bf p},\omega)d\omega/2\pi=\gamma^0$ for the spectral function. This equation is satisfied approximately when the value of $\Gamma$ is small: $\Gamma\lesssim M$. If this expression is substituted into the trace in Eq.(14), it is written as
\begin{equation}
 {\rm Tr}[\rho\gamma^2\rho\gamma^2]=\frac{16N_{c}N_{f}}{X^2}(8M^2 p_y^2-X)\Gamma^2.
\end{equation}%
Here $N_{c}$ and $N_{f}$ denote the numbers of the color and flavor of the quark and the use is made of the definition: $X\equiv (p^2-M^2+\Gamma^2)^2+4M^2\Gamma^2$. Substituting this equation into Eq.(14), the shear viscosity is written as
\begin{equation}
\eta = \frac{64N_{c}N_{f}}{T}\int\frac{\mathrm{d}\varepsilon}{2\pi}
 \int\frac{\mathrm{d}{\bf p}}{(2\pi)^3} \frac{{p_x ^2}{p_y ^2}M^2 \Gamma^2 n(\varepsilon)(1-n(\varepsilon))}{[(p^2-M^2+\Gamma^2)^2+4M^2\Gamma^2]^2}.
\end{equation}%
where we have neglected the second term in Eq.(17) because the main contribution in the above integral comes from the integral region where $X$ is small.

Now let us discuss the results of the numerical calculations of the shear viscosity. Since we are interested in the QGP phase, our calculations are restricted to the case of high temperature or high density. As for the effective quark mass, we take $M=100{\rm MeV}$ ($N_{c}=3$ and $N_{f}=2$) for convenience. In Fig.3, the shear viscosity is drawn as a function of the width $\Gamma$ at $T=150{\rm MeV}$ and $T=200{\rm MeV}$ in the case of $\mu=10{\rm MeV}$.
\begin{figure}
 \includegraphics[width=\linewidth]{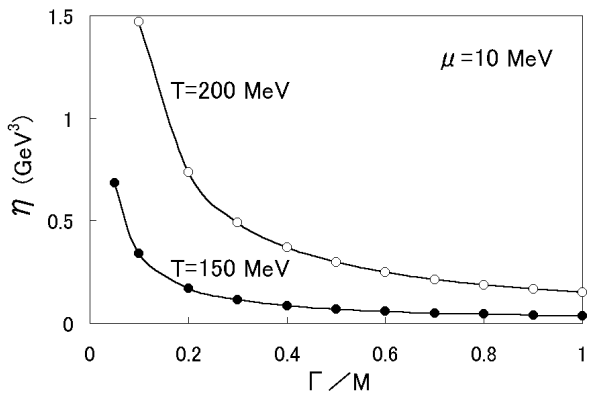}
 \caption{\label{fig:epsart}The shear viscosity as a function of the width $\Gamma$ at the temperature $T=150{\rm MeV}$ (brack circles) and $T=200{\rm MeV}$ (white circles) in the case of $\mu=10{\rm MeV}$.}
\end{figure}
Similarly Fig.4 shows the shear viscosity as a function of $\Gamma$ at $\mu=10{\rm MeV}$ and $\mu=100{\rm MeV}$ in the case of $T=150{\rm MeV}$.
\begin{figure}
 \includegraphics[width=\linewidth]{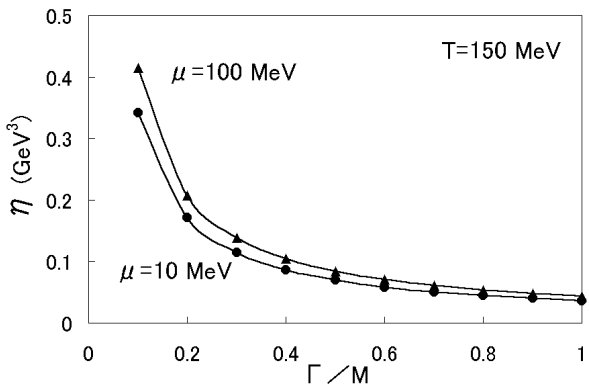}
 \caption{\label{fig:epsart}The shear viscosity as a function of width $\Gamma$ at the chemical potential $\mu=10{\rm MeV}$ (black circles) and $\mu=100{\rm MeV}$ (triangles) in the case of $T=150{\rm MeV}$.}
\end{figure}
These figures evidently show that the viscosity is a rapidly decreasing function of the width $\Gamma$. The viscosity seems to diverge at $\Gamma=0$ where the quark matter becomes ideal gas. This is understood by noting that $(\Gamma/(x^2+\Gamma^2))^2\rightarrow (\pi\delta(x))^2$ as $\Gamma\rightarrow 0$. As the $\Gamma$ becomes larger, the viscosity decreases rapidly. This is also apparent because $(\Gamma/(x^2+\Gamma^2))^2\rightarrow 0$ as $\Gamma\rightarrow \infty$. The width means the inverse of the quark life time and reflects the strength of the interaction. Therefore the quark matter becomes almost perfect fluid under strongly interacting state. This property is consistent with the recent RHIC data. But we do not discuss the numerical values of the shear viscosity in comparison with the experimental data. The reason is that the spectral function used in this paper is a phenomenological one and has not been determined theoretically.

We also comment the dependence of the shear viscosity on the temperature and the density. From Figs. 3 and 4, it is seen that the viscosity increases as the temperature increases. As for the density, it is a slowly increasing function. These dependences seem to be consistent with the classical expression of the viscosity of gas: $\eta=\rho vl/3$ ($\rho$=density, $v$=velocity and $l$= mean free path of the particle).

In conclusion, we have calculated the shear viscosity of the quark matter with the use of the Kubo-Mori formula. We have obtained the shear viscosity expressed in terms of the quark spectral function. Assuming a modified Bright-Wigner type of the spectral function, the shear viscosity is a rapidly decreasing function of the width; the quark matter approaches to perfect fluid under strongly interacting state.

\begin{acknowledgments}
The authors would like to thank Professors S. Sakai (Yamagata University), M.Asakawa (Osaka University) and K.Iida (Kochi University) for invaluable comments. They also thank the Yukawa Institute for Theoretical Physics at Kyoto University. Discussions during the YITP workshop YITP-W-04-07 on Thermal Quantum Field Theories and Their Applications were very useful to start and develop this work. 
\end{acknowledgments}

\end{document}